\documentclass[acp, manuscript]{copernicus}

\usepackage{xcolor}

\begin{document}
\nolinenumbers

\title{Predicting Gas-Particle Partitioning Coefficients of Atmospheric Molecules with Machine Learning}


\Author[1]{Emma}{Lumiaro}
\Author[1]{Milica}{Todorovi{\'c}}
\Author[2]{Theo}{Kurten}
\Author[3]{Hanna}{Vehkam{\"a}ki}
\Author[1]{Patrick}{Rinke}

\affil[1]{Department of Applied Physics, Aalto University, P.O. Box 11100, 00076 Aalto, Espoo, Finland}
\affil[2]{Department of Chemistry, Faculty of Science, PO Box 55, FI-00014 University of Helsinki, Finland}
\affil[3]{Institute for Atmospheric and Earth System Research/Physics, Faculty of Science, PO Box 64, FI-00014 University of Helsinki, Finland}



\correspondence{Patrick Rinke (patrick.rinke@aalto.fi)}

\runningtitle{TEXT}

\runningauthor{TEXT}

\received{}
\pubdiscuss{} 
\revised{}
\accepted{}
\published{}


\firstpage{1}

\maketitle

\begin{abstract}
The formation, properties and lifetime of secondary organic aerosols in the atmosphere are largely determined by gas-particle partitioning coefficients of the participating organic vapours. Since these coefficients are often difficult to measure and to compute, we developed a machine learning model to predict them given molecular structure as input. Our data-driven approach is based on the dataset by Wang et al. (Atmos. Chem. Phys., 17, 7529 (2017)), who computed the partitioning coefficients and saturation vapour pressures of 3414 atmospheric oxidation products from the master chemical mechanism using the COSMOtherm program. We train a kernel ridge regression (KRR) machine learning model on the saturation vapour pressure ($\mathrm{P}_{\mathrm{sat}}$), and on two equilibrium partitioning coefficients: between a water-insoluble organic matter phase and the gas phase ($\mathrm{K}_{\mathrm{WIOM/G}}$), and between an infinitely dilute solution with  pure water  and the gas phase ($\mathrm{K}_{\mathrm{W/G}}$). For the input representation of the atomic structure of each organic molecule to the machine, we test different descriptors. We find that the many-body tensor representation (MBTR) works best for our application, but the topological fingerprint (TopFP) approach is almost as good, and is significantly more cost effective.
Our best machine learning model (KRR  with Gaussian kernels+ MBTR) predicts $\mathrm{P}_{\mathrm{sat}}$ and $\mathrm{K}_{\mathrm{WIOM/G}}$ to within 0.3 logarithmic units and $\mathrm{K}_{\mathrm{W/G}}$ to within 0.4 logarithmic units of the original COSMOtherm calculations. This is equal or better than the typical accuracy of COSMOtherm predictions compared to experimental data (where available). We then apply our machine learning model to a dataset of 35,383 molecules  that we generated based on a carbon 10 backbone and functionalized with 0 to 6 carboxyl (-COOH), carbonyl (=O) or hydroxyl (-OH) functional groups to evaluate its performance for polyfunctional compounds with potentially  low saturation vapour pressures $\mathrm{P}_{\mathrm{sat}}$. The resulting saturation vapor pressure and partitioning coefficient distributions were physico-chemically reasonable, and the volatility predictions for the most highly oxidized compounds were in qualitative agreement with experimentally inferred volatilities of atmospheric oxidation products with similar elemental composition. 
\end{abstract}


\introduction  

Aerosols in the atmosphere are fine solid or liquid  particles (or droplets) suspended in air. They scatter and absorb solar radiation and form cloud droplets in the atmosphere, affect visibility and human health and are responsible for large uncertainties in the study of climate change. Most aerosol particles  are secondary organic aerosols (SOAs) that are formed by oxidation of volatile organic compounds (VOCs), which are in turn emitted into the atmosphere for example from plants or traffic. Some of the oxidation products  have volatilities low enough to condense.  The formation, growth and lifetime of SOAs is governed largely by the concentrations, saturation vapour pressures ($P_{\mathrm{sat}}$) and equilibrium partitioning coefficients of the participating vapours. While real atmospheric aerosol particles are extremely complex mixtures of very many different organic and inorganic compounds \citep{Elm/etal:2020}, partitioning of organic vapours is by necessity usually modelled in terms of a few representative parameters. These include the saturation vapour pressure, describing the interaction of a compound with itself, and various partitioning coefficients ($\mathrm{K}$), describing the interaction of the compound with a few representative other species. Typical partitioning coefficients in chemistry include ($\mathrm{K}_{\mathrm{W/G}}$) for the partitioning between the gas phase and pure water (i.e. an infinitely dilute solution of the compound), and ($\mathrm{K}_{\mathrm{O/W}}$) for the partitioning between octanol and water solutions \footnote{The gas-octanol partitioning coefficient ($\mathrm{K}_{\mathrm{O/G}}$) can then be obtained from these by division.}. For describing organic aerosols, the partitioning coefficient between the gas phase and a model water-insoluble organic matter phase (WIOM; $\mathrm{K}_{\mathrm{WIOM/G}}$) is likely better than ($\mathrm{K}_{\mathrm{O/G}}$).

Unfortunately, measuring any of these partitioning coefficients is challenging, especially for multifunctional low-volatility compounds most relevant to SOA formation, and little experimental data is thus available for the atmospherically most interesting organic vapour species. For relatively simple organic compounds, efficient empirical parametrizations have been developed to predict their condensation-relevant properties. These include poly-parameter linear free-energy relationships (ppLFERs) \citep{Goss/Schwarzenbach:2001,Goss:2004,Goss:2006}, GROup contribution Method for Henry’s law Estimate (GROMHE) \citep{Raventos-Duran/etal:2010}, and SPARC Performs Automated Reasoning in Chemistry (SPARC)  \citep{Hilal/etal:2008}, SIMPOL \citep{Pankow/Asher:2008}, EVAPORATION \citep{Compernolle/etal:2011}, and Nannoolal \citep{Nannoolal/etal:2008}. Many of these parameterisations are available in a user-friendly format on the UManSysProp website \citep{Topping/etal:2016}. However, due to the limitations in the available experimental datasets on which they are based, the accuracy of such approaches typically degrades significantly once the compound contains more than three or four functional groups \citep{Valorso/etal:2011}.

Approaches based on quantum chemistry such as COSMO-RS (COnductor-like Screening MOdel for Real Solvents, \cite{Klamt/Eckert:2000,Klamt/Eckert:2003,Eckert/Klamt:2002}), implemented for example in the COSMOtherm program, can calculate saturation vapour pressures and partitioning coefficients also for complex polyfunctional compounds, albeit only with order-of-magnitude accuracy. However, for many applications even this is extremely useful. For example in the context of new-particle formation (often called nucleation), knowing if the saturation vapour pressure of an organic compound is above or below about $10^{-14}$ atm is vital, as only species with a $P_{\mathrm{sat}}$ lower than this can be assumed to condense irreversibly onto  preexisting nanometer-sized cluster. Even lower $P_{\mathrm{sat}}$ is required for the vapour to form completely new particles. This illustrates the challenge in performing experiments on SOA-relevant species: a compound with a saturation vapour pressure of e.g. $10^{-10}$ atm at room temperature would be considered non-volatile in terms of most available measurement methods - but yet its volatility is far too high to allow nucleation in the atmosphere.

COSMO-RS/COSMOtherm calculations are based on density functional theory (DFT), and within the context of quantum chemistry they are thus considered computationally reasonably affordable compared to high-level ab initio methods such as coupled cluster theory. Nevertheless, the application of COSMO-RS to complex polyfunctional organic molecules still requires significant computational effort, especially due to the conformational complexity of these species, for example in terms of different potential hydrogen bonding patterns. Overall, there could be up to $10^4-10^7$ different organic compounds in the atmosphere (not even counting most oxidation intermediates), which makes the computation of saturation vapour pressures and partitioning coefficients a daunting task \citep{Shrivastava/etal:2019, Ye/etal:2016}.

Here, we take a different approach compared to previous parametrization studies, and consider a data-science perspective \citep{Himanen/Geurts/Foster/Rinke:2019}. Instead of assuming chemical or physical relations, we let the data speak for itself. We develop and train a machine learning model to extract patterns from available data and predict saturation vapour pressures as well as partitioning coefficients.

Machine learning has only recently spread into atmospheric science \citep{Cervone/etal:2008,Toms/etal:2018,Barnes/etal:2019,Nourani/etal:2019,Huntingford/etal:2019,Masuda/etal:2019}. Prominent applications include the identification of forced climate patterns \citep{Barnes/etal:2019}, precipitation prediction \citep{Nourani/etal:2019}, climate analysis \citep{Huntingford/etal:2019}, pattern discovery \citep{Toms/etal:2018}, risk assessment of atmospheric emissions \citep{Cervone/etal:2008}, and the estimation of cloud optical thicknesses \citep{Masuda/etal:2019}. In molecular and materials science, machine learning is more established and now frequently complements theoretical or experimental methods \citep{Rampi_review, ma_deep_2015,shandiz_application_2016, gomez_design_2016, bartok_machine_2017,rlb2018,goldsmith_machine,meyer_machine_2018,Zunger:2018,Gu/etal:2019,Schmidt/etal:2019,Jensen/etal:2020,Coley/etal:2020}. Here we build on our experience in atomistic, molecular machine learning \citep{Gosh/etal:2019,Todorovic/etal:2019,Stuke/etal:2019,Himanen/etal:2020} to train a regression model that maps molecular structures onto saturation vapour pressures and partitioning coefficients. Once trained, the machine learning model can make  saturation vapour pressure and partitioning predictions at COSMOtherm accuracy for hundreds of thousands of new molecules at no further computational cost. When experimental training data becomes available, the machine learning model can easily be extended to make predictions for experimental pressures and coefficients.

Due to the above-mentioned lack of comprehensive experimental databases for saturation vapour pressures or gas-liquid partioning coefficient of polyfunctional atmospherically relevant molecules, our machine-learning model is based on the computational data by \cite{Wang/etal:2017}. They computed the partitioning coefficients and  saturation vapour pressures for 3414 atmospheric secondary oxidation products, obtained from the Master Chemical Mechanism \citep{Jenkin/etal:1997,Saunders/etal:2003}, using a combination of quantum chemistry and statistical thermodynamics as implemented in the COSMOtherm approach \citep{Klamt/Eckert:2000}.

We transform the molecular structures in Wang's dataset into atomistic descriptors more suitable for machine learning than the atomic coordinates or the commonly used simplified molecular-input line-entry system (SMILES) strings. Optimal descriptor choices have been the subject of increased research in recent years \citep{Langer/Goessmann/Rupp:2020,Rossi/Cumby:2020,Himanen/etal:2020}. We here test several descriptor choices: the many body tensor representation \citep{Huo/Rupp:2017}, the Coulomb matrix \citep{Rupp/etal:2012}, the Molecular ACCess System (MACCS) structural key \citep{Durant/etal:2002}, a topological fingerprint developed by RDkit \citep{landrum2006rdkit} based on the daylight fingerprint \citep{james1995daylight} and the Morgan fingerprint \citep{Morgan:1965}. 

In this work, we address the following objectives. 1) With view to future machine learning applications in atmospheric science, we assess the predictive capability of different structural descriptors for machine learning the chosen target properties. 2) We quantify the predictive power of our KRR machine learning model for Wang's dataset to ascertain if the dataset size is sufficient for accurate machine learning predictions. 3) We then apply our validated machine learning model to a new molecular dataset to gain chemical insight into SOA condensation processes.

The paper is organized as follows. We describe our machine learning methodology in section \ref{sec:method}, then present the  machine learning results in section \ref{sec:results}. Section \ref{sec:pred} demonstrates how we employed the trained model for fast prediction of molecular properties. We discuss our findings and present a summary in section \ref{sec:conc}.

\section{Methods}
\label{sec:method}

\begin{figure}[h]
    \centerline{\includegraphics[width=0.99\columnwidth]{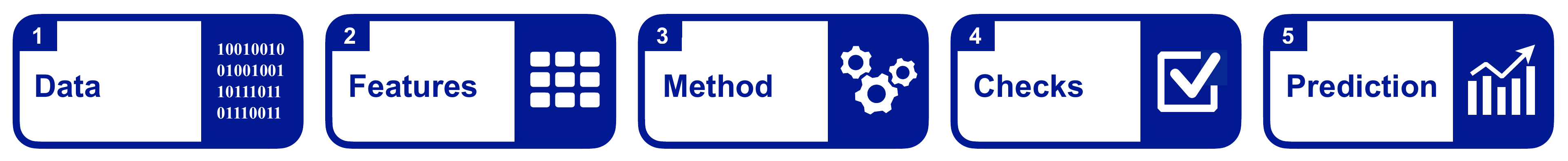}}
    \caption{Schematic representation of our machine learning workflow}
\label{fig:MLapproach}
\end{figure}

Our machine learning approach has five components as illustrated in Fig.~\ref{fig:MLapproach}. We start off with the raw data, which we present and analyse in section~\ref{sec:dataset}. The raw data is then transformed into a suitable representation for machine learning (step 2). We introduce five different representations in section~\ref{sec:features}, which we test in our machine learning model (cf section~\ref{sec:results}). Next we choose our machine learning method. Here we use kernel ridge regression (KRR), which is introduced in section~\ref{sec:ML}. We analyse the learning success of our machine learning approach in step 4. The results of this process are shown in section~\ref{sec:results}. In this step we also make adjustments to the representation and the parameters of the model to improve the learning. Finally, we use the best machine learning model to make predictions as shown in section~\ref{sec:pred}.

\subsection{Dataset}
\label{sec:dataset}

In this work we are interested in the the equilibrium partitioning coefficients of a molecule between a water-insoluble organic matter (WIOM) phase  and  gas phase ($\mathrm{K}_{\mathrm{WIOM/G}}$) as well as  gas phase and infinitely dilute water solution. 
These coefficients are defined as 
\begin{align}
	& K_\mathrm{WIOM/G}  = \displaystyle{ \frac{C_\mathrm{WIOM}}{C_\mathrm{G}}} \\
	& K_\mathrm{W/G}  = \displaystyle{\frac{C_\mathrm{W}}{C_\mathrm{G}} },
\end{align}
where $C_{\mathrm{WIOM}}$, $C_\mathrm{W}$, and $C_\mathrm{G}$ are the equilibrium concentrations of the molecule in the WIOM, water, and gas phase, respectively, at the limit of infinite dilution. In the framework of COSMOtherm calculations, gas-liquid partitioning coefficients can be converted into saturation vapor pressures, or vice versa, using the activity coefficients $\gamma_\mathrm{W}$ or $\gamma_\mathrm{WIOM}$  in the corresponding liquid (which can also be computed by COSMOtherm). Specifically, if for example $K_\mathrm{W/G}$ is expressed in units of $m^3g^{-1}$, then $P_\mathrm{sat} = \frac{RT}{M \gamma_\mathrm{W} K_\mathrm{W}}$, where $R$ is the gas constant, $T$ the temperature, $M$ the molar mass of the compound and $K_\mathrm{W}$ and $\gamma_\mathrm{W}$ are the partitioning and activity coefficients in water \citep{Arp/Goss:2009}. We caution, however, that many different conventions exist e.g. for the dimensions of the partitioning coefficients, as well as the reference states for activity coefficients -- the relation given above applies only for the particular conventions used by COSMOtherm.

\cite{Wang/etal:2017} used the conductor-like screening model for real solvents (COSMO-RS) theory \citep{Klamt/Eckert:2000} implemented in COSMOtherm to calculate the two partitioning coefficients $\mathrm{K}_{\mathrm{WIOM/G}}$\footnote{As a model WIOM phase, Wang \textit{et al.} used a compound originally suggested by \cite{Kalberer/etal:2004} as a representative secondary organic aerosol constituent. The IUPAC name for the compound in question, with elemental composition C$_{14}$H$_{16}$O$_5$, is 1-(5-(3,5-dimethylphenyl)dihydro-[1,3]dioxolo[4,5-d][1,3]dioxol-2-yl)ethan-1-one.} and $\mathrm{K}_{\mathrm{W/G}}$ for 3414 molecules. These molecules were generated from 143 parent volatile organic compounds with the Master Chemical Mechanism (MCM) \citep{Jenkin/etal:1997,Saunders/etal:2003} through photolysis and reactions with ozone, hydroxide and nitrade. 

\begin{figure}[h]
    \centerline{\includegraphics[width=0.99\columnwidth]{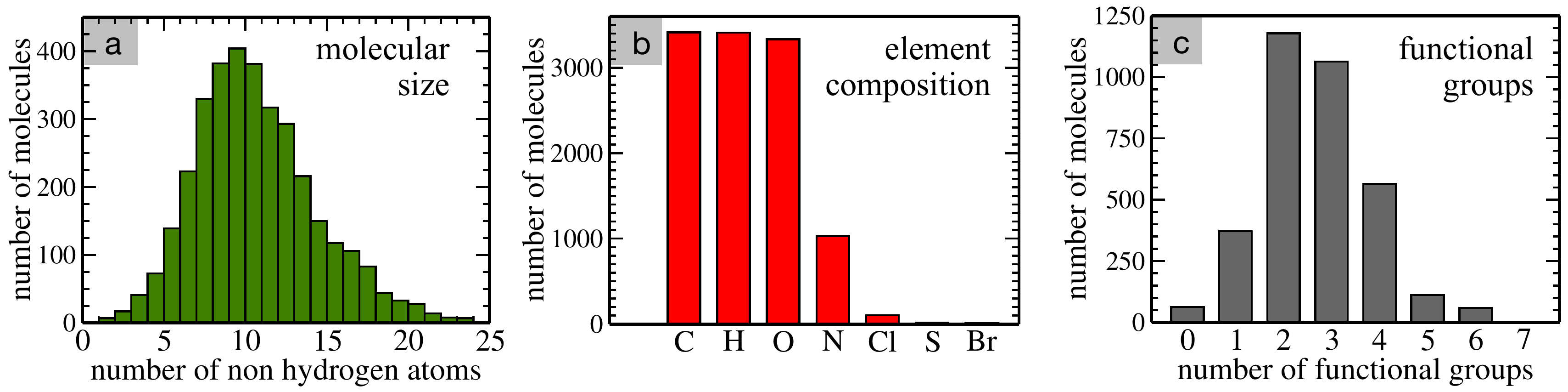}}
    \caption{Dataset statistics: Panel a) shows the size distribution (in terms of the number of non-hydrogen atoms) of all 3414 molecules in the dataset.  Panel b) illustrates how many molecules contain each of the chemical species and panel c) depicts the functional group distribution. }
\label{fig:datasetmol}
\end{figure}

Here, we analyse the composition of the publicly available dataset by Wang \textit{et al.} in preparation for machine learning. Figure~\ref{fig:datasetmol} illustrates key dataset statistics. Panel a) shows the size distribution of  molecules as measured in number of atoms. The 3414 non-radical species obtained from MGM range in size from 4 to 48 atoms, which translates into 1 to 24 non-hydrogen atoms per molecule. The distribution peaks at 10 non-hydrogen atoms and is skewed towards larger molecules. 
Panel b) illustrates how many molecules contain at least one atom of the indicated element. All molecules contain carbon (100\% C), 3410 contain hydrogen (H; 99.88\%) and 3333 also oxygen (O; 97.63\%). Nitrogen (N) is the next most abundant element (30.17\%) followed by chlorine (Cl; 3.05\%), sulphur (S; 0.44\%) and bromide (Br; 0.32\%).  Lastly, panel c) presents the distribution of functional groups. It peaks at 2 to 3 functional groups per molecule, with relatively few molecules having 0, 5 or 6 functional groups.

\begin{figure}[h]
    \centerline{\includegraphics[width=0.99\columnwidth]{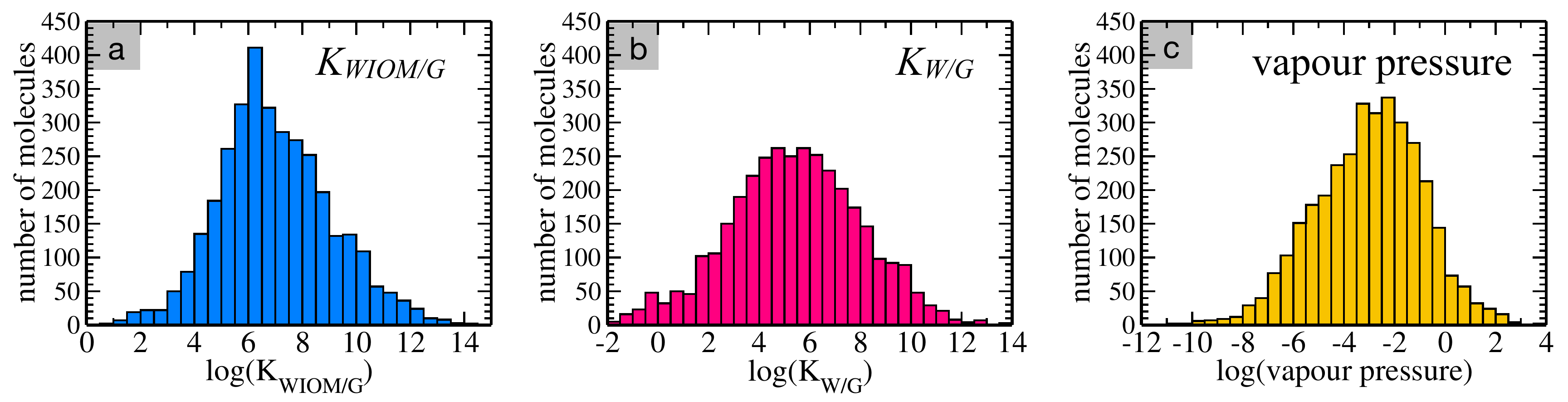}}
    \caption{Dataset statistics: distributions of equilibrium partitioning coefficients a) $K_{WIOM/G}$, b) $K_{W/G}$ and c) the saturation vapour pressure $\mathrm{P}_{\mathrm{sat}}$ for all 3414 molecules in the dataset.}
\label{fig:datasetcoeffs}
\end{figure}

Figure~\ref{fig:datasetcoeffs} shows the distribution of the target properties $\mathrm{K}_{\mathrm{WIOM/G}}$, $\mathrm{K}_{\mathrm{W/G}}$ and $\mathrm{P}_{\mathrm{sat}}$ in Wang's dataset on a logarithmic scale. The equilibrium partitioning coefficient $\mathrm{K}_{\mathrm{WIOM/G}}$ distribution is skewed slightly towards larger coefficients, in contrast to the saturation vapour pressure  $\mathrm{P}_{\mathrm{sat}}$ distribution with an asymmetry towards molecules with lower pressures. All three target properties cover approximately 15 logarithmic units and are approximately Gaussian distributed. Such peaked distributions are often not ideal for machine learning since they over-represent molecules near the peak of the distribution and under-represent molecules at their edges. The data peak does supply enough similarity to ensure good quality learning, but properties of the under-represented molecular types might be harder to learn.

Wang's dataset of 3414 molecules is relatively small for machine learning, which often requires hundreds of thousands to millions of training samples \citep{Pyzer-Knapp/etal:2015,Smith/Isayev/Roitberg:2017,Stuke/etal:2019,Gosh/etal:2019}. A slightly larger set of Henry's law constants, which are related to $\mathrm{K}_{\mathrm{W/G}}$, were reported  by \cite{Sander:2015} for 4632 organic species. Sander's database is a collection of 17350  Henry's law constant values collected from 689 references and therefore not as internally consistent as Wang's dataset. We are not aware of a larger dataset that reports partitioning coefficients. For this reason, we rely exclusively on Wang's dataset and show that we can develop machine learning methods that are just as accurate as the underlying calculations and thus suitable for predictions.

\subsection{Representations}
\label{sec:features}

The molecular representation for machine learning should fulfil certain requirements. It should be invariant with respect to translation and rotation of the molecule and permutations of atomic indices. Furthermore, it should be continuous, unique, compact and efficient to compute \citep{Faber/etal:2015,Huo/Rupp:2017,Langer/Goessmann/Rupp:2020,Himanen/etal:2020}. 

In this work we employ two classes of representations for the molecular structure, also  known as descriptors:  \emph{physical} and \emph{cheminformatics} descriptors. \emph{Physical descriptors} encode physical distances and angles between atoms in the material or molecule. Such descriptors generally exhibit good performance for many different system types. Meanwhile, decades of research in \emph{cheminformatics} have produced topological descriptors that encode the qualitative aspects of molecules in a compact representation. These descriptors are typically bitvectors, in which molecular features are encoded (hashed) into binary fingerprints, which are joined into long binary vectors. In this work, we use two physical descriptors, the Coulomb Matrix and the many-body tensor, and three cheminformatics descriptors: the MACCS structural key, the topological fingerprint and the Morgan fingerprint. 

In Wang's dataset the molecular structure is encoded in SMILES (Simplified Molecular Input Line Entry Specification) strings. We convert these SMILES strings into structural descriptors using Open Babel \citep{OBoyle/etal:2011} and the DScribe library  \citep{Himanen/etal:2020} or into cheminformatics descriptors using RDkit  \citep{landrum2006rdkit}. 

\subsubsection{Coulomb Matrix} 

The Coulomb matrix CM descriptor is inspired by an electrostatic representation of a molecule \citep{Rupp/etal:2012}. It encodes the cartesian coordinates of a molecule in a simple matrix of the form

\begin{equation}
  C_{ij} = 
  \begin{cases}
    0.5Z_i^{2.4} & \text{if $i=j$}\\
    \frac{Z_i Z_j}{\lVert \mathbf{R}_i-\mathbf{R}_j \rVert } & \text{if $i \neq j$}\\
  \end{cases} 
\end{equation}
where $\mathbf{R}_i$ is the coordinate of atom $i$ with atomic charge $Z_i$. The diagonal provides element-specific information. The coefficient and the exponent have been fitted to the total energies of isolated atoms \citep{Rupp/etal:2012}. Off-diagonal elements encode inverse distances between the atoms of the molecule by means of a Coulomb-repulsion-like term. 

\begin{figure}[h]
    \centerline{\includegraphics[width=0.99\columnwidth]{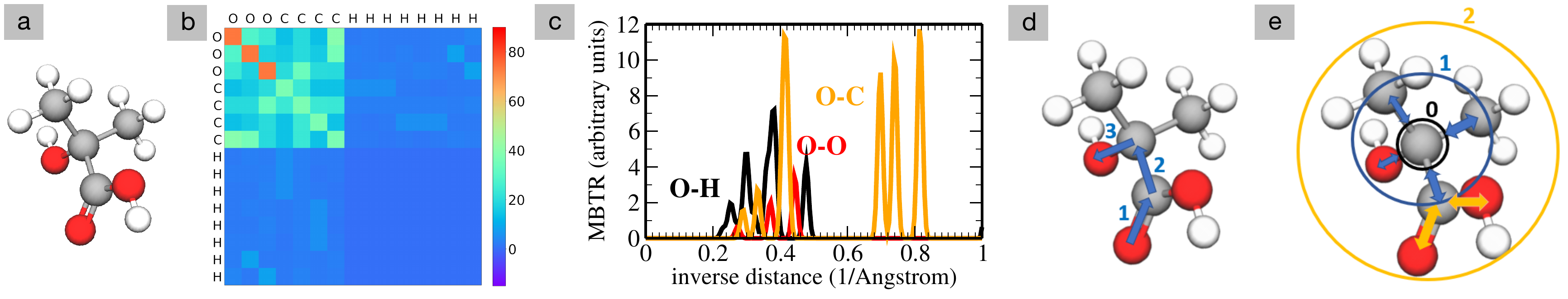}}
    \caption{Pictorial overview over descriptors used in this work: a) ball and stick model of 2-hydroxy-2-methylpropanoic acid, b) corresponding Coulomb matrix (CM), c) the O-H, O-O and O-C inverse distance entries of the many-body tensor representation (MBTR), d) topological fingerprint (TopFP) depiction of a path with length three, and e) Morgan circular fingerprint with radius 0 (black), radius 1 (blue) and radius 2 (orange).}
\label{fig:desc}
\end{figure}

The dimension of the Coulomb matrix is chosen to fit the largest molecule in the data set, i.e. it corresponds to the number of atoms of the largest molecule. The ``empty'' rows of Coulomb matrices for smaller molecules are padded with zeroes. Invariance with respect to the permutation of atoms in the molecule is enforced by simultaneously sorting rows and columns of each Coulomb matrix in descending order according to their $\ell^2$-norms. An example of a Coulomb matrix for 2-hydroxy-2-methylpropanoic acid is shown in Fig.~\ref{fig:desc}b.

The CM is easily understandable, simple and relatively small as a descriptor. However, it performs best with Laplacian kernels in the machine-learning model (see Section~\ref{sec:ML}), while other descriptors work better with the more standard choice of a Gaussian kernel.

\subsubsection{Many-body tensor representation} 

The many-body tensor representation (MBTR) follows the Coulomb matrix philosophy of encoding the internal coordinates of a molecule. We will here describe the MBTR only qualitatively. Detailed equations can be found in the original publication \citep{Huo/Rupp:2017}, our previous work \citep{Himanen/etal:2020,Stuke/etal:2020} or  Appendix~\ref{sec:AppendixMBTR}. 

Unlike the Coulomb matrix, the many-body tensor is continuous and it distinguishes between different types of internal coordinates. At many-body level 1, the MBTR records the presence of all atomic species in a molecule by placing a Gaussian at the atomic number on an axis from 1 to the number of elements in the periodic table. The weight of the Gaussian is equal to the number of times the species is present in the molecule. At many-body level 2, inverse distances between every pair of atoms (bonded and non-bonded) are recorded in the same fashion. Many-body level 3 adds angular information between any triple of atoms. Higher levels (e.g. dihedral angles) would in principle be straightforward to add, but are not implemented in the current MBTR versions \citep{Huo/Rupp:2017,Himanen/etal:2020}. Figure~\ref{fig:desc}c shows selected MBTR elements for 2-hydroxy-2-methylpropanoic acid. 

The MBTR is a continuous descriptor, which is advantageous for machine learning. However, MBTR is by far the largest descriptor out of the five we tested, and this can pose restrictions on memory and computational cost. Furthermore, the MBTR is more difficult to interpret than the CM.

\subsubsection{MACCS Structural Key}

The Molecular ACCess System (MACCS) structural key is a dictionary-based descriptor \citep{Durant/etal:2002}. It is represented as a bitvector of Boolean values that encode answers to a set of predefined questions. The MACCS structural key we used is a 166 bit long set of answers to 166 questions such as "Is there a S-S bond" or "Does it contain Iodine?" \citep{landrum2006rdkit, james1995daylight}.

MACCS is the smallest out of the five descriptors and extremely fast to use. Its accuracy critically depends on how well the 166 questions encapsulate the chemical detail of the molecules. Is it likely to reach a moderate accuracy with low computational cost and memory usage and could be beneficial for fast testing of a machine learning model.

\subsubsection{Topological Fingerprint}

The topological fingerprint (TopFP) is RDKit's original fingerprint \citep{landrum2006rdkit} inspired by the Daylight fingerprint \citep{james1995daylight}. TopFP  hashes the atomic structure. This means that the structure is divided into smaller substructures and each substructure is converted into a unique binary ID called a hash. The hashes are concatenated into a long bitvector representing the entire molecule. In TopFP, the molecule is hashed along topological paths, or along bonds, as illustrated in Figure \ref{fig:desc}d. The path starts from one atom in a molecule and travels along bonds until $k$ bond lengths have been traversed, and that completes a hash. The length of the bitvector, maximum and minimum possible path lengths $k_{max}$ and $k_{min}$ and the length of one hash can be optimized. 


Topology is an informative molecular feature. We therefore expect TopFP to balance good accuracy with reasonable computational cost. However, this binary fingerprint is difficult to visualize and analyse for chemical insight.

\subsubsection{Morgan Fingerprint}

The Morgan fingerprint is also a bit-vector constructed by hashing the molecular structure. In contrast to the Topological fingerprint, the Morgan fingerprint is hashed along circular or spherical paths around the central atom as illustrated in Figure \ref{fig:desc}e. Each substructure for a hash is constructed by first numbering the atoms in a molecule with unique integers by applying the Morgan algorithm. Each uniquely numbered atom then becomes a cluster center, around which we iteratively increase a spherical radius to include the neighbouring bonded atoms \citep{rogers2010extended}. Each radius increment extends the neighbour list by another molecular bond. The ``circular'' substructures found by the algorithm described above, excluding duplicates, are then hashed into a fingerprint \citep{james1995daylight, landrum2006rdkit}. The length of the fingerprint and the maximum radius can be optimized. 

The Morgan fingerprint is quite similar to the TopFP in size and type of information encoded, so we expect similar performance. It does not lend itself to easy chemical interpretation.

\subsection{Machine Learning Method}
\label{sec:ML}

\subsubsection{Kernel Ridge Regression}

In this work, we apply the kernel ridge regression (KRR) machine learning method. KRR is an example of supervised learning, in which the machine learning model is trained on pairs of input ($x$) and target ($f$) data. The trained model then predicts target values for previously unseen inputs. In this work, the input $x$ are the molecular descriptors CM and MBTR as well as the MACCS, TopFP and Morgan fingerprints. The targets are scalar values for the equilibrium partitioning coefficients and saturation vapour pressures. 

KRR is based on Ridge Regression, in which a penalty for overfitting is added to an ordinary least squares fit \citep{friedman2001elements}. In KRR, unlike Ridge regression, a nonlinear kernel is applied. This maps the molecular structure to our target properties in a high dimensional space \citep{Stuke/etal:2019,rupp2015machine}.

The target values $f$ are a linear expansion in kernel elements
\begin{align} \label{eq:KRR}
    f(x) = \sum^n_{i=1}\alpha_ik(x_i,x),
\end{align}
where the sum runs over all training molecules. In this work, we use two different kernels, the  Gaussian kernel 
\begin{align} \label{eq:ga}
    k_{G}(x, x') =e^{-\gamma\|x-x'\|_2^2}
\end{align}
and the Laplacian kernel  
\begin{align} \label{eq:la}
    k_{L}(x, x') = e^{-\gamma\|x-x'\|_1} .
\end{align}
The kernel width $\gamma$  is a hyperparameter of the KRR model.

The regression coefficients $\alpha_i$ can be solved by minimizing the error 
\begin{align}\label{eq:error} 
    \min_{\alpha} \sum^n_{i=1}(f(x_i)-y_i)^2 + \lambda\pmb{\alpha}^T\mathbf{K}\pmb{\alpha},
\end{align}
where $y_i$ are reference target values for molecules in the training data. The second term is the regularization term, whose size is controlled by the  hyperparameter $\lambda$. $\mathbf{K}$ is the kernel matrix of training inputs $k(x_i,x_j)$.

This minimization problem can be solved analytically for the expansion coefficients $\alpha_i$ 

\begin{align}\label{eq:alpha}
    \pmb{\alpha} = (\bf{K} - \lambda \bf{I})^{-1}\bf{y}
\end{align}
The hyperparameters $\gamma$ and $\lambda$ need to be optimised separately.

We implemented KRR in Python using \emph{scikit-learn}  \citep{pedregosa2011scikit}. Our implementation has been described in Ref. \cite{Stuke/etal:2019,Stuke/etal:2020}

\subsubsection{Computational Execution}
\label{sec:CE}

Data used for supervised machine learning is typically divided into two sets, a large training set and a small test set. Both sets consists of input vectors and corresponding target properties. The training set is used to train the KRR model, while the test set molecules are unseen by the trained model. The test set thus quantifies the model performance. At the outset, we separate a test set of 414 molecules. From the remaining molecules, we choose six different training sets of size 500, 1000, 1500, 2000, 2500 and 3000, so that a smaller training size is always a subset of the larger one. Training the model on a sequence of such training sets allows us to compute a \emph{learning curve}, which facilitates the assessment of learning success with increasing training data size. We quantify the accuracy of our KRR model by computing the mean absolute error (MAE) for the test set. To get statistically meaningful results, we repeat the training procedure 10 times. In each run, we shuffle the dataset before selecting the training and test sets so that the KRR model is trained and tested on different data each time. Each point on the learning curves is computed as the average over 10 results, and the spread serves as the standard deviation of the datapoint.

Model training proceeds by computing the KRR regression coefficients $\alpha_i$, obtained by minimizing equation \ref{eq:error}. KRR hyperparameters $\gamma$ and $\lambda$ are typically optimized via grid search, and average optimal solutions are obtained by cross-validating the procedure. In cross-validation we split off a validation set from the training data before training the KRR model. KRR is then trained for  all possible combinations of discretised hyperparameters (grid search) and evaluated on the validation set.  This is done several times, so that the molecules in the validation set are changed each time. Then the hyperparameter combination with minimum average cross-validation error is chosen. Our implementation of cross-validated grid search is also based on Scikit-learn \citep{pedregosa2011scikit}.

\begin{table}[H]
\centering
\caption{All the hyperparameters that were optimized.}\label{table:hyptable}
\begin{tabular}{|l|l|l|}
\hline
& Hyperparameters   & Optimized Values \\  \hline
KRR   & width of the kernel $\gamma$,  regularization parameter $\lambda$     & descriptor-dependent \\ \hline
MBTR  & broadening parameters $\sigma_2, \sigma_3$; weighting parameters $w_2,w_3$  & 0.0075, 0.1; 1.2, 0.8    \\ \hline
TopFP & vector length;  maximum path length $k_{max}$  ;   bits per hash  & 8192; 8; 16   \\ \hline
Morgan &  vector length; radius;  & 2048; 2  \\ \hline
\end{tabular}
\end{table}

Table \ref{table:hyptable} summarises all the hyperparameters optimised in this study, those for KRR and the molecular descriptors, and their optimal values. In grid search, we varied both $\gamma$ and $\lambda$ by ten values between $10^{-1}$ and $10^{10}$. In addition, we used two different kernels, Laplacian and Gaussian. We compared the performance of the two kernels for the average of 5 runs for each training size and the most optimal kernel was chosen. In cases in which both kernels performed equally well, e.g., for the fingerprints, we chose the Gaussian kernel for its lower computational cost. 

MBTR hyperparameters and TopFP hyperparameters were optimized by grid search for several training set sizes (MBTR for sizes 500, 1500 and 3000 and TopFP for size 1000 and 1500 ) and the average of two runs for each training size was taken. We did not extend the descriptor hyperparameter search to larger training set sizes, since we found that the hyperparameters were insensitive to the training set size. The MBTR weighting parameters were optimized in 8 steps between 0 (no weighting) and 1.4, and the broadening parameters in 6 steps between $10^{-1}$ and $10^{-6}$. The length of TopFP was varied between 1024 and 8192 (size can varied by $2^{n}$). The range for the maximum path length extended from 5 to 11 and the bits per hash were varied between 3 and 16.

\section{Results}
\label{sec:results}

\begin{figure}[h]
    \centerline{\includegraphics[width=0.99\columnwidth]{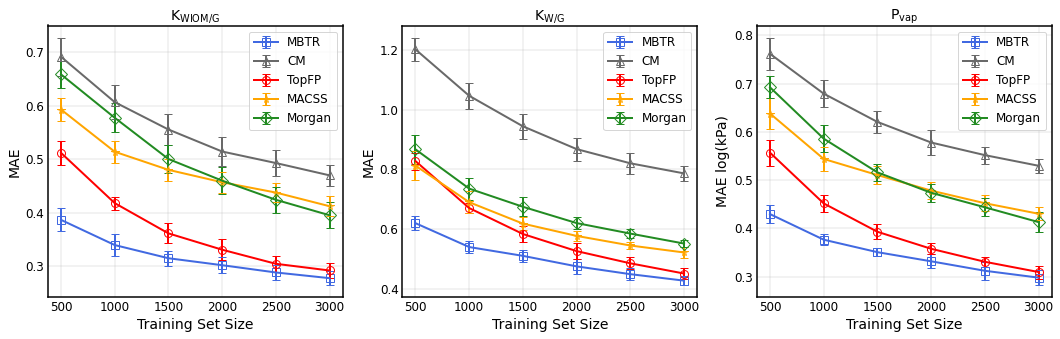}}
    \caption{The learning curves for equilibrium partitioning coefficients  $K_\mathrm{WIOM/G}$,  $K_\mathrm{W/G}$ and saturation vapour pressure $P_\mathrm{sat}$ for predictions made with all five descriptors.}
\label{fig:learning_curves}
\end{figure}

In Figure \ref{fig:learning_curves} we present the learning curves for our objectives $K_\mathrm{WIOM/G}$,  $K_\mathrm{W/G}$ and  $P_\mathrm{sat}$. Shown is the mean average error (MAE) as a function of the training set size for all three target properties and with all five molecular descriptors. As expected, the MAE decreases as the training size increases. For all target properties, the lowest errors are achieved with MBTR and the worst performing descriptor is CM. TopFP approaches the accuracy of MBTR as the training size increases and appears likely to outperform MBTR beyond the largest training size of 3000 molecules. 

Table \ref{tab:3000_table} summarises the average MAEs and their standard deviations for the best-trained KRR model (training size of 3000 with MBTR descriptor). The highest accuracy is obtained for partitioning coefficient $K_\mathrm{WIOM/G}$, with a  mean average error of 0.278, i.e. only 1.9\% of the entire $K_\mathrm{WIOM/G}$ range. The second best accuracy is obtained for saturation vapour pressure  $P_\mathrm{sat}$ with an MAE of 0.298 (or 2.0\% of the range of pressure values). The lowest accuracy is obtained for $K_\mathrm{W/G}$ with an MAE of 0.428. However, the range for partitioning coefficient $K_\mathrm{W/G}$ is also the largest, as seen in Figure \ref{fig:learning_curves}, so this amounts to only 2.7\% of the entire range of values. Our best machine learning MAEs are of the order of the COSMOtherm prediction accuracy, which lies at around a few tenths of log values \citep{Stenzel/Goss/Endo:2014,Schroeder/Fulem/Martins:2016,vanderSpoel/etal:2019}.





\begin{figure}[h]
    \centerline{\includegraphics[width=0.99\columnwidth]{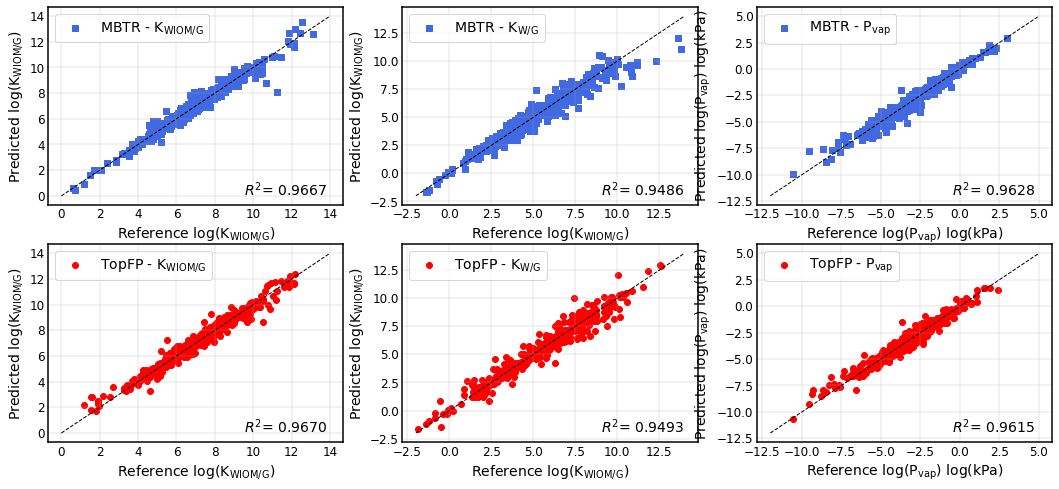}}
    \caption{The scatter plots for predictions for partitioning coefficients of a molecule between a water-insoluble organic matter and gas phase $K_\mathrm{WIOM/G}$, water and gas phase $K_\mathrm{W/G}$ and  the saturation vapour pressure $P_\mathrm{sat}$ for the test set of 414 molecules using MBTR (top) and TopFP (bottom). The prediction with the lowest mean average error was chosen for each scatter plot.}
\label{fig:scatter_plots}
\end{figure}

Figure \ref{fig:scatter_plots} shows the results for the best-performing descriptors MBTR and TopFP in more detail. The scatter plots illustrate how well the KRR predictions match the reference values. The match is further quantified by $\mathrm{R}^2$ values. For all three target values, the predictions hug the diagonal quite closely and we observe only a few outliers that are further away from the diagonal. The predictions of  partitioning coefficient $K_\mathrm{WIOM/G}$ are most accurate. This is expected because the MAE in Table \ref{tab:3000_table} is lowest for this property. The largest scattered is observed for partitioning coefficient $K_\mathrm{W/G}$ which had the highest MAE in Table \ref{tab:3000_table}.

\begin{table}[H]
\centering
\caption{The average mean average errors (MAE) and the standard deviations for all the descriptors and target properties (equilibrium partitioning coefficients  $K_\mathrm{WIOM/G}$,  $K_\mathrm{W/G}$ and saturation vapour pressure $P_\mathrm{sat}$) with the largest possible training size of 3000. }
\begin{tabular}{|l|l|l|l|l|l|l|}
\hline
            & \multicolumn{2}{l|}{$K_\mathrm{WIOM/G}$} & \multicolumn{2}{l|}{$K_\mathrm{W/G}$} & \multicolumn{2}{l|}{$P_\mathrm{sat}$} \\ \hline
Descriptor  & MAE            & $\Delta$               & MAE           & $\Delta$              & MAE log(kPa)    & $\Delta$log(kPa)   \\ \hline
CM          & 0.470          & ± 0.020         & 0.787         & ± 0.028        & 0.530           & ± 0.016     \\ \hline
MBTR        & 0.278          & ± 0.013         & 0.427         & ± 0.015        & 0.298           & ± 0.016     \\ \hline
MACCS       & 0.407          & ± 0.018         & 0.520         & ± 0.020        & 0.428           & ± 0.016     \\ \hline
Morgan      & 0.408          & ± 0.030         & 0.549         & ± 0.018        & 0.403           & ± 0.012     \\ \hline
TopFP & 0.292          & ± 0.030         & 0.453         & ± 0.026        & 0.307           & ± 0.017     \\ \hline
\end{tabular}
\label{tab:3000_table}
\end{table}

\section{Predictions}
\label{sec:pred}

In the previous section we showed that our KRR model trained on the Wang \textit{et al.} dataset produces low prediction errors for molecular partitioning coefficients and can now be employed as a fast predictor. When shown further molecular structures, it can make instant predictions for the molecular properties of interest. We demonstrate this application potential on an example dataset generated to imitate organic molecules typically found in the atmosphere.

Atmospheric oxidation reaction mechanisms can be generally classified into two main types: fragmentation and functionalization. For SOA formation, functionalization is more relevant, as it leads to products with intact carbon backbones and added polar (and volatility-lowering) functional groups. Many of the most interesting molecules from a SOA-forming point of view, e.g. monoterpenes, have around 10 carbon atoms. These compounds simultaneously have high enough emissions or concentrations to produce appreciable amounts of condensable products, while being large enough for those products to have low volatility. 

\begin{figure}[h]
    \centerline{\includegraphics[width=0.99\columnwidth]{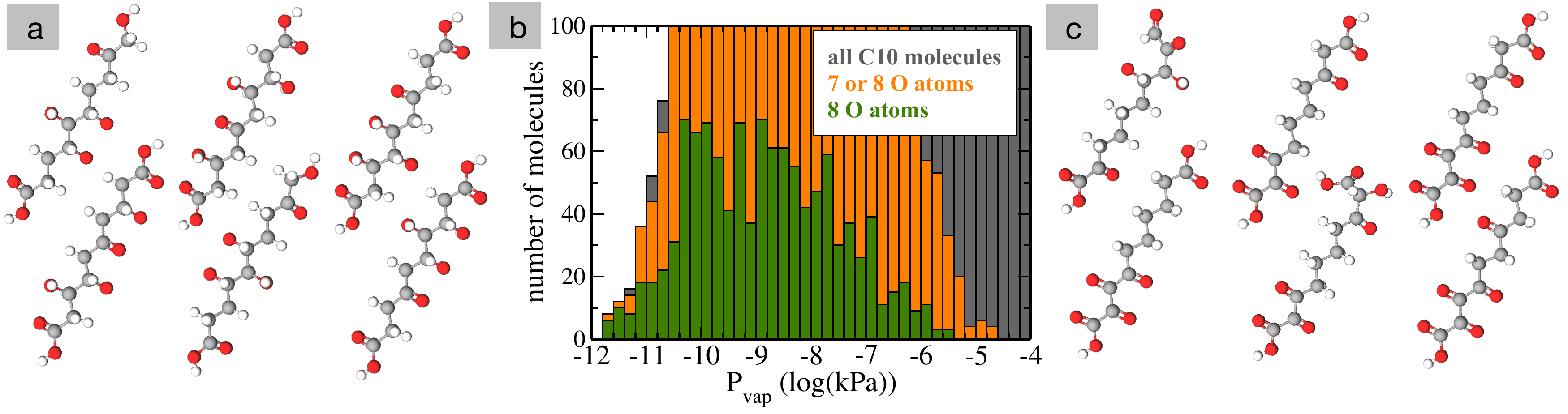}}
    \caption{a) Atomic structure of the 6 molecules with the lowest predicted saturation vapour pressure $P_{\mathrm{sat}}$; b) $P_{\mathrm{sat}}$ histograms for molecules containing 7 or 8 O atoms (orange) or only 8 O atoms (green). For reference, the histogram of all molecules (grey) is also shown. c) Atomic structure of the 3 molecules with 7 and 8 O atoms and the highest saturation vapour pressure $P_{\mathrm{sat}}$.}
\label{fig:C10Vap78}
\end{figure}

We thus generate a dataset of molecules with a backbone of ten carbon (C10) atoms. For simplicity, we use a linear alkane chain. In analogy with Wang's dataset, we then decorate this backbone with 0 to 6 functional groups at different locations. We limit ourselves to the typical groups formed in "functionalizing" oxidation of VOC by both of the main day-time oxidants OH and O$_3$: carboxyl(-COOH), carbonyl (=O) and hydroxyl (-OH) \citep{Seinfeld/Spyros:2016}. The (-COOH) group can only be added to the ends of the C10 molecule, while (=O) and (-OH) can be added to any carbon atom in the chain. We then generate all possible combinations combinatorially and filter out duplicates resulting from symmetric combinations of functional groups. In total we obtain 35,383 unique molecules. Example molecules are depicted in Figure \ref{fig:C10Vap78}. 

\begin{figure}[h]
    \centerline{\includegraphics[width=0.99\columnwidth]{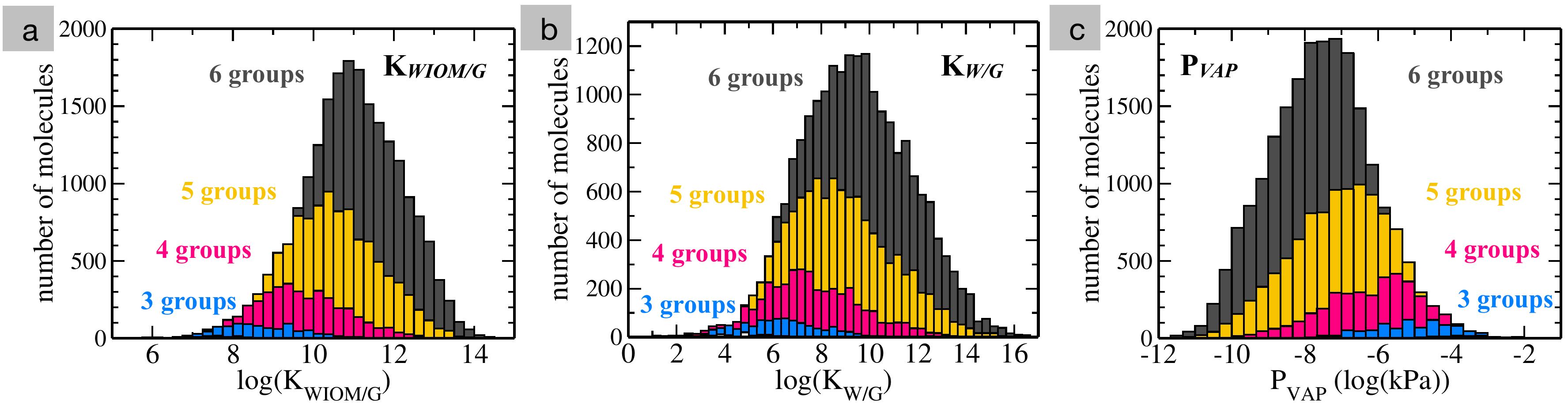}}
    \caption{Histograms of C10 TopFP-KRR predictions for a) $K_\mathrm{WIOM/G}$,b ) $K_{W/G}$ and c) $P_{\mathrm{sat}}$. The histograms are divided into different numbers of functional groups. Molecules with 2 or fewer functional groups have been omitted from these histograms, because their total number is very low in the C10 dataset.}
\label{fig:C10Pred}
\end{figure}

For each of the 35,383 molecules we generated a SMILES string that serves as input for the TopFP fingerprint. We did not relax the geometry of the molecules with force fields or density-functional theory. We then predicted $P_{\mathrm{sat}}$, $K_{\mathrm{WIOM/G}}$ and $K_{\mathrm{W/G}}$ with the TopFP-KRR model. We chose TopFP as descriptor, because its accuracy is close to that of the best performing MBTR KRR model, but significantly cheaper to evaluate. 

\begin{figure}[h]
    \centerline{\includegraphics[width=0.99\columnwidth]{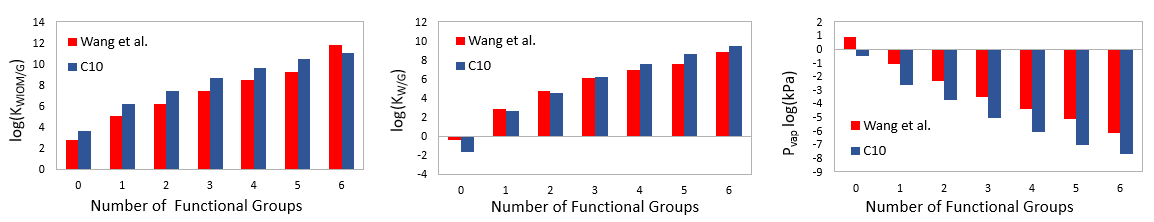}}
    \caption{Histograms visualizing the averages of  equilibrium partitioning coefficients $K_\mathrm{WIOM/G}$,  $K_\mathrm{W/G}$ and   saturation vapour pressure $P_\mathrm{sat}$ for different numbers of functional groups for our  C10 dataset (in blue) and Wang's dataset (in red).}
\label{fig:hist_means}
\end{figure}

Figures~\ref{fig:C10Pred} and \ref{fig:hist_means} show the predictions of our TopFP-KRR model for the C10 dataset. For comparison with Wang's dataset, we broke the histograms and analysis down into the number of functional groups. The comparison between our C10 and Wang's dataset in Figure~\ref{fig:hist_means} shows that the averages of all three quantities agree well for different numbers of functional groups, which illustrates that both datasets are similar. A certain degree of similarity is required to ensure predictive power, since machine learning models do not extrapolate well to data that lies outside the training range.

Figure~\ref{fig:C10Pred} illustrates that the saturation vapour pressure $P_{\mathrm{sat}}$ decreases with increasing number of functional groups as expected, whereas $K_\mathrm{WIOM/G}$ and $K_\mathrm{W/G}$ increase. This is consistent with Wang's dataset as shown in Fig.~\ref{fig:hist_means}, where we compare averages between the two datasets. The magnitude of the decrease (increase) amounts to approximately 1 or 2 orders of magnitude per functional group and is consistent with existing structure-activity relationships based on experimental data (e.g. \cite{Pankow/Asher:2008,Compernolle/etal:2011,Nannoolal/etal:2008}).

The region of low $P_{\mathrm{sat}}$ is most relevant for atmospheric SOA formation. Figure~\ref{fig:C10Vap78}b shows histograms of only molecules with 7 or 8 oxygen atoms. These are compared to the full dataset. Since the ``8 O atom set'' is a subset of the ``7 or 8 O atoms'' set, which in turn is a subset of ``all molecules'' the length of the bars in a given bin reflect the percentages of molecules with 7 or 8 O atoms. We observe that below 10$^{-10}$ kPa, almost all C10 molecules contain 7 or 8 O atoms, as there is little grey visible in that part of the histogram. In the context of atmospheric chemistry, the least-volatile fraction of our C10 dataset corresponds to LVOC ("low volatility organic compounds"), which are capable of condensing onto small aerosol particles, but not actually forming them. Our results are thus in qualitative agreement with recent experimental results by \citep{Peraekylae/etal:2020}, who concluded that the highly oxidized C$_{10}$ products of $\alpha$-pinene oxidation are mostly LVOC. However, we note that the compounds measure by Per\"akyl\"a et al. are likely to contain functional groups not included in our C10 dataset, as well as structural features such as branching and rings.  Figure~\ref{fig:C10Vap78}a and Figure~\ref{fig:C10Vap78}c show the molecular structures of the lowest-volatility compounds, as well as the highest-volatility compounds with 7 or 8 O atoms, respectively. (Note that the latter set inevitably contains at least one carboxylic acid group, as we have restricted the number of functional groups to six or less, and only the acid groups contain two oxygen atoms.) Comparing the two sets, we can see that the lowest-volatility compounds contain more hydroxyl groups, and less ketone groups, while the highest-volatility compounds with 7 or 8 oxygen atoms contain almost no hydroxyl groups. This is expected - for example according to the SIMPOL model \citep{Pankow/Asher:2008}, a hydroxyl group lowers the saturation vapor pressure by over a factor of 100 at 298 K, while the effect of a ketone group is a bit less than a factor of 10. However, even the lowest-volatility compounds (Figure~\ref{fig:C10Vap78}a) contain a few ketone groups, such that the number of hydrogen-bond donor and acceptor groups are roughly similar. This result demonstrates that unlike the simplest group-contribution models, both the original COSMOTherm predictions, and the machine-learning model based on them, are capable of accounting for hydrogen-bonding interactions between functional groups.



\section{Conclusions}
\label{sec:conc}

 
In this study, we set out to evaluate the potential of the KRR machine learning method to map molecular structures to its atmospheric partitioning behaviour, and establish which molecular descriptor has the best predictive capability. 

KRR is a relatively simple kernel-based machine-learning technique that is straightforward to implement and fast to train. Given model simplicity, the quality of learning depends strongly on information content of the molecular descriptor. More specifically, it hinges on how well each format encapsulates the structural features relevant to the atmospheric behaviour. The exhaustive approach of MBTR descriptor to documenting molecular features has led to very good predictive accuracy in machine learning of molecular properties \citep{Stuke/etal:2019,Langer/Goessmann/Rupp:2020,Rossi/Cumby:2020,Himanen/etal:2020} and this work is no exception. The lightweight CM descriptor does not perform nearly as well, but these two representations from physical sciences provide us with an upper and lower limit on predictive accuracy. 

Descriptors from cheminformatics that were developed specifically for molecules have variable performance. Between them, the topological fingerprint leads to best learning quality that approaches MBTR accuracy in the limit of larger training set sizes. This is a notable finding, not least because the relatively small TopFP data structures in comparison to MBTR reduce the computational time and memory required for machine learning. MBTR encoding requires knowledge of the 3-dimensional molecular structure, which raises the issue of conformer search. It is unclear which molecular conformers are relevant for atmospheric condensation behaviour, and COSMOterm calculations on different conformers can produce values that are orders of magnitude apart. TopFP requires only connectivity information and can be built from SMILES strings, eliminating any conformer considerations (albeit at the cost of possibly losing some information on e.g. intramolecular hydrogen bonds). All this makes TopFP the most promising descriptor for future machine learning studies in atmospheric science that we have identified in this work.

Our results show that KRR can be used to train a model to predict COSMOtherm saturation vapor pressures, with error margins smaller than those of the original COSMOtherm predictions. In the future, we propose to extend our training set to encompass especially atmospheric autoxidation products \citep{Bianchi/etal:2019}, which are not included in existing saturation vapour pressure datasets, and for which existing prediction methods are highly uncertain. While COSMOtherm predictions for such molecules also have large uncertainties, a fast and efficient "COSMOtherm - level" KRR predictor would still be immensely useful, for example in evaluating whether a given compound is likely to have extremely low volatility, or not. As experimental data for such compounds becomes available, either through indirect inference methods such as \cite{Peraekylae/etal:2020} or for example thermal desorption methods \citep{Li/etal:2020}. These could then be used to constrain and anchor the model, and ultimately yield also quantitatively reliable volatility predictions. 






\appendix

\section{Many-body tensor representation}    
\label{sec:AppendixMBTR}


In this appendix we provide the mathematical structure of the MBTR as it is implemented in the DScribe library \cite{Himanen/etal:2020}. The many-body levels in the MBTR are denoted $k$. For $k=1, 2, 3$, geometry functions encode the different features: $g_1(Z_{l})=Z_{l}$ (atomic number), $g_2(\boldsymbol{R}_{l}, \boldsymbol{R}_{m})=|\boldsymbol{R}_{l}- \boldsymbol{R}_{m}|$ (distance) or $g_2(\boldsymbol{R}_{l}, \boldsymbol{R}_{m})=\frac{1}{|\boldsymbol{R}_{l}- \boldsymbol{R}_{m}|}$ (inverse distance), and $g_3(\boldsymbol{R}_{l}, \boldsymbol{R}_{m}, \boldsymbol{R}_{n})=\textnormal{cos}(\angle(\boldsymbol{R}_{l}- \boldsymbol{R}_{m}, \boldsymbol{R}_{n}- \boldsymbol{R}_{m}))$ (cosine of angle).

The scalar values returned by the geometry functions $g_k$ are Gaussian broadened into continuous representations $\mathcal{D}_k$:
\begin{equation}
\mathcal{D}_{1}^{l}(x)=\frac{1}{\sigma_1 \sqrt{2 \pi}}e^{-\frac{(x-g_1(Z_l))^2}{2\sigma_1^2}}
\end{equation}

\begin{equation}
\mathcal{D}_{2}^{l,m}(x)=\frac{1}{\sigma_2 \sqrt{2 \pi}}e^{-\frac{(x-g_2(\boldsymbol{R}_l, \boldsymbol{R}_m))^2}{2\sigma_2^2}}
\end{equation}
\begin{equation}
\mathcal{D}_{3}^{l,m,n}(x)=\frac{1}{\sigma_3 \sqrt{2 \pi}}e^{-\frac{(x-g_3(\boldsymbol{R}_l, \boldsymbol{R}_m), \boldsymbol{R}_n)^2}{2\sigma_3^2}}.
\end{equation}
The $\sigma_k$'s are the feature widths for the different $k$-levels and $x$ runs over a predefined range $[x_\textrm{min}^k,x_\textrm{max}^k]$ of possible values for the geometry functions $g_k$. 

Finally, a weighted sum of distributions $\mathcal{D}_k$ is generated for each possible combination of chemical elements present in the dataset
\begin{equation}
\textnormal{MBTR}_{1}^{Z_1}(x) = \sum_{l}^{|Z_1|}w_1^l\mathcal{D}
_1^l(x)
\end{equation}

\begin{equation}
\textnormal{MBTR}_{2}^{Z_1, Z_2}(x) = \sum_{l}^{|Z_1|} \sum_{m}^{|Z_2|}w_2^{l,m}\mathcal{D}
_2^{l,m}(x)
\end{equation}
\begin{equation}
\textnormal{MBTR}_{3}^{Z_1, Z_2, Z_3}(x) = \sum_{l}^{|Z_1|} \sum_{m}^{|Z_2|} \sum_{n}^{|Z_3|}w_3^{l,m,n}\mathcal{D}
_3^{l,m,n}(x).
\end{equation}
The sums for $l$, $m$, and $n$ run over all atoms with atomic numbers $Z_1$, $Z_2$ and $Z_3$. $w_k$ are weighting functions that balance the relative importance of different $k$-terms and/or limit the range of inter-atomic interactions. For $k=1$, usually no weighting is used ($w_1^l=1$). For $k=2$ and $k=3$ the following exponential decay functions are implemented in \texttt{DScribe}
\begin{equation}
w_2^{l,m} = e^{-s_k|\boldsymbol{R}_{l}- \boldsymbol{R}_{m}|}
\end{equation}
\begin{equation}
w_3^{l,m,n} = e^{-s_k(|\boldsymbol{R}_{l}- \boldsymbol{R}_{m}|+|\boldsymbol{R}_{m}- \boldsymbol{R}_{n}|+|\boldsymbol{R}_{l}- \boldsymbol{R}_{n}|)}
\end{equation}
The parameter $s_k$ effectively tunes the cutoff distance. The functions MBTR$_k(x)$ are then discretized with $n_k$ many points in the respective intervals $[x_\textrm{min}^k,x_\textrm{max}^k]$.

\noappendix       




\appendixfigures  

\appendixtables   





\begin{acknowledgements}
This work was supported by the Academy of Finland (project number 316601) and through their Flagship programme: Finnish Center for Artificial Intelligence FCAI. This work was further supported by the European Research Council project 692891-DAMOCLES, by COST (European Cooperation in Science and Technology) Action 18234 and by the University of Helsinki Faculty of Science ATMATH project. We thank CSC, the Finnish IT Center for Science  and Aalto Science IT for computational resources. 
\end{acknowledgements}








\bibliographystyle{copernicus}
\bibliography{references.bib}

\end{document}